\documentstyle[12pt]{article}
\def\be{\begin{equation}}
\def\ee{\end{equation}}
\def\bear{\be\begin{array}}
\def\eear{\end{array}\ee}
\def\bea{\begin{eqnarray}}
\def\eea{\end{eqnarray}}

\def\baselinestretch{1}
\begin{document}
\catcode`@=11
\newtoks\@stequation
\def\subequations{\refstepcounter{equation}%
\edef\@savedequation{\the\c@equation}%
  \@stequation=\expandafter{\theequation}
  \edef\@savedtheequation{\the\@stequation}
  \edef\oldtheequation{\theequation}%
  \setcounter{equation}{0}%
  \def\theequation{\oldtheequation\alph{equation}}}
\def\endsubequations{\setcounter{equation}{\@savedequation}%
  \@stequation=\expandafter{\@savedtheequation}%
  \edef\theequation{\the\@stequation}\global\@ignoretrue

\noindent}
\catcode`@=12
\begin{titlepage}
\title{{\bf Orbifolds with continuous Wilson lines and soft
terms}
}
\author{{\bf H.B. Kim}
\\
{\small Departamento de F\'{\i}sica Te\'orica C--XI} \\
{\small Universidad Aut\'onoma de Madrid, Cantoblanco, 28049 Madrid, Spain}\\
{\small hbkim@delta.ft.uam.es}
\\
\hspace{3cm}\\
{\bf C. Mu\~noz}\thanks{On leave of absence from Departamento de
F\'{\i}sica Te\'orica C-XI, Universidad Autonoma de Madrid, 
Cantoblanco, 28049 Madrid,
Spain.}
\\ 
{\small Department of Physics} \\
{\small Korea Advanced Institute of Science and Technology,
        Taejon 305-701, Korea} \\
{\small cmunoz@bosque.sdi.uam.es}}

\date{}
\maketitle
\def\baselinestretch{1.15}
\begin{abstract}
\noindent

Orbifold compactifications with continuous Wilson lines have very
interesting 
characteristics and as a consequence they are 
candidates to obtain realistic models. 
We perform an analysis of the soft supersymmetry-breaking terms arising
in this type of compactifications.
We also 
compare these results with those of orbifolds without 
including continuous Wilson lines. 
Their phenomenological properties turn out to be
similar. 

\end{abstract}

\thispagestyle{empty}

\leftline{}
\leftline{}
\leftline{FTUAM 96/41}
\leftline{KAIST-TH-96/18}
\leftline{December 1996}

\vskip-20.cm
\rightline{FTUAM 96/41}
\rightline{KAIST-TH-96/18}
\vskip3in

\end{titlepage}
\newpage
\setcounter{page}{1}


\noindent The compactification of the $E_8\times E_8$ heterotic string on 
orbifolds \cite{orbifolds} proved in the past to be a helpful way 
to construct
semirealistic models. In particular, when quantized 
Wilson lines \cite{ibanez1}
were included in the analysis of orbifold compactifications, 
smaller gauge groups
containing the standard model 
with appropriate matter representations were obtained \cite{ibanez2}.
Although the rank of these models is still 16 and the matter 
content too large, both
can be reduced due to the effect of Fayet-Iliopoulos terms associated
to anomalous $U(1)'s$ \cite{standard}.

On the other hand, there is an alternative 
mechanism to produce the breaking
of the gauge symmetry obtained from orbifold compactification. 
This can be
carried out through the use of continuous Wilson line background 
fields \cite{cwl}. 
They are untwisted moduli fields (in addition to the usual
$T_i$ and $U_i$ moduli) whose vacuum expectation values (VEVs) 
break the gauge
group lowering the rank.
In addition they give mass to some matter fields. This is 
therefore a potential mechanism to obtain relevant models.
Another interesting phenomenological property of continuous Wilson lines
is the following. The discrepancy between the unification scale of gauge
couplings and the string unification scale may be explained by the effect
of string threshold corrections \cite{unification}. However, 
in order to achieve it,
the VEVs of the $T$ moduli must be quite large, ${\rm Re}\,T\simeq 16$,
far away from the self-dual point. 
The advantage of using
continuous Wilson line moduli is that 
the same threshold effects may be obtained but
with moderate values of the moduli \cite{Nilles}.

As it is well known, in order to connect the above models with the
low-energy phenomena, the computation of the soft 
supersymmetry (SUSY)-breaking
terms is crucial \cite{Munoz1}. 
This was carried out for the case of (0,2) symmetric Abelian orbifolds,
just assuming that the seed of SUSY breaking is located in the
dilaton/moduli sectors \cite{IDL,Brignole,Ferrara,Brignole2},
with interesting results.
For example, non-universal scalar masses, even for particles with the
same overall modular weight, are allowed and 
certain general sum-rules involving soft terms of different
particles are obtained \cite{Brignole2}. 
However, the case of orbifolds with
continuous Wilson lines has not been addressed in the literature yet. 
Recently, due to the interesting
phenomenological properties of the latter, there has been some
activity in trying to obtain information about the structure of 
their low-energy
effective Lagrangian \cite{kahler,gauge}. 
In particular, the explicit dependence of the
K\"ahler potential on the Wilson line moduli has been 
obtained in \cite{kahler}
allowing the computation of the soft 
terms in this type of models. This is
the aim of the present letter.  
We will ask what changes with respect to the previous analyses if one
allows the presence of continuous Wilson lines.

The work of \cite{kahler} was mainly focus on 
the so-called factorizable {\boldmath $Z_N$}
orbifolds for which the 
underlying six-dimensional torus lattice {\boldmath $T_6$}
can be decomposed into a direct sum of a four-dimensional and a
two-dimensional sublattice {\boldmath $T_4\oplus T_2$}.
Then, the following K\"ahler
potentials for the untwisted moduli spaces associated with the torus
{\boldmath $T_2$} were obtained
\begin{eqnarray}
K &=& -\log(T+T^*-AA^*)\ ,
\label{kahl1}
\\
K &=& -\log[(T+T^*)(U+U^*)-(B+C^*)(B^*+C)]\ ,
\label{kahl2}
\end{eqnarray}
where $A$, $B$, $C$ are the complex Wilson line fields. The K\"ahler
potential (\ref{kahl2}) is obtained in orbifold models where the complex
plane associated with {\boldmath $T_2$} has order two.
In this case $T$-, $U$-type
moduli and two Wilson lines $B$ and $C$ are present and the 
K\"ahler potential
possesses the required
structure to solve the $\mu$ problem \cite{GM}. Let us concentrate first
on case (\ref{kahl1}). Then, the K\"ahler potential associated
with the three complex planes, including matter fields $C_i$,
$i=1,2,3$, is 
given by
\begin{eqnarray}
K &=& -\sum_{i=1,2} \log(T_i+T_i^*-C_iC_i^*)-
\log(T_3+T_3^*-AA^*-C_3C_3^*)\
\nonumber\\ &&
\simeq {\hat K}(T_i,T_i^*,A,A^*) + {\tilde K}_i(T_i,T_i^*,A,A^*) C_iC_i^*
\label{kahl3}
\end{eqnarray}
with
\begin{eqnarray}
{\tilde K}_i &=& (T_i+T_i^*)^{-1}\ ,\ i=1,2
\hbox{\ \ ;\ \ }
{\tilde K}_3 = (T_3+T_3^*-AA^*)^{-1}\ ,
\label{kahl4}
\\
{\hat K} &=& \sum_i {\hat K}_i 
\hbox{\ \ ,\ \ }
{\hat K}_i=\log{\tilde K}_i\ ,
\label{kahl5}
\end{eqnarray}
where we associate the Wilson line $A$ with (say) the third complex plane.
If the moduli space contains more than one Wilson line modulus 
the above equations will be the same with the modification
$AA^*\rightarrow \sum_k A_kA_k^*$.
Notice that when continuous Wilson lines are turned off, the usual 
untwisted
K\"ahler potential 
is recovered.

Although the origin of SUSY breaking in strings is by itself an 
outstanding problem, we can get information about the SUSY-breaking
sector of the theory parameterizing our ignorance by the VEVs of the
dilaton and moduli auxiliary fields \cite{Brignole,Brignole2}.
Since the moduli metric is off-diagonal due to the $T_3$-$A$ 
mixing (\ref{kahl4},\ref{kahl5}), 
we will use for our calculation the parameterization
introduced in \cite{CY} where the off-diagonal case was analyzed in
order 
to compute Calabi-Yau soft terms. Then 
\begin{eqnarray}
F_S &=& \sqrt3m_{3/2}\sin\theta (S+S^*)e^{-i\gamma_S}\ , \nonumber\\
F_\alpha &=& \sqrt3m_{3/2}\cos\theta (P\Theta)_{\alpha}\ ,
\label{parametrization}
\end{eqnarray}
where the angle $\theta $ and the complex parameters 
$\Theta _{\alpha}$ just parameterize the
direction of the Goldstino in the $S$, $T_i$, $A$ field space.
The analysis when more than one Wilson line modulus is present
is straightforward and will be carried out below.
$P$ is a matrix canonically
normalizing the moduli 
fields \cite{CY}, i.e.  
$P^\dagger\hat{K}P=1$ where $1$ stands for the unit matrix
and 
${\hat K}_{\alpha \beta}\equiv {\bar {\partial}}_{\alpha}
{\partial }_{\beta} {\hat K}$.
Imposing 
$\sum_{\alpha} \Theta_{\alpha}^*\Theta_{\alpha}=1$
this parametrization has the virtue that,
when we plug it in the general form of the supergravity scalar potential,
the cosmological constant
\begin{eqnarray}
V_0 &=& (S+S^*)^{-2}|F_S|^2+F_{\alpha}^{\dagger}
{\hat K}_{\alpha \beta}F_{\beta}-3m_{3/2}^2 
\label{conscosmo}
\end{eqnarray}
vanishes by
construction. 
If the value of $V_0$ is not assumed to be zero
the modifications are straightforward \cite{Brignole,Brignole2}.

It is obvious, that the main difference with the previous 
computations of soft terms in orbifold compactifications is that now
one must add in the VEV of the scalar potential a contribution 
due to the auxiliary field associated with the Wilson line modulus.
Thus the soft terms are given
by
\begin{eqnarray}
m_i^2 &=& m_{3/2}^2\left[ 1
-3\cos^2\theta(P\Theta)_{\alpha}^{\dagger}
 \left({\bar \partial}_{\alpha}
 \partial_{\beta} \log {\tilde K}_i
\right)
(P\Theta)_{\beta} \right]\ ,
\label{scalar.mass.square}
\\
A_{123} &=& -\sqrt3m_{3/2}\left[
\sin\theta e^{-i\gamma_S}
+\cos\theta (P\Theta)_{\alpha}
\left\{\sum_{j=1,2,3}
\partial_{\alpha} \log {\tilde K}_j
-\partial_{\alpha} \hat{K}-\partial_{\alpha} \log Y_{123}\right\}
\right]\ .
\label{trilinear}
\end{eqnarray}
The above scalar masses and trilinear scalar couplings correspond to
matter fields which have already been canonically normalized. Here
$Y_{123}$ is a renormalizable Yukawa coupling involving three
untwisted fields and $A_{123}$ is its corresponding trilinear soft term.
 
Due to property (\ref{kahl5}) the soft masses can now easily be computed
since 
${\bar {\partial}}_{\alpha}
{\partial }_{\beta}\log {\tilde K}_i
= ({\hat K}_i)_{\alpha \beta}$. Recalling that
$P^\dagger\hat{K}P=1$ we obtain 
\begin{eqnarray}
m_1^2 &=& m_{3/2}^2 (1-3\cos^2\theta |\Theta_1|^2)\ ,
\label{soft masses1} 
\\
m_2^2 &=& m_{3/2}^2 (1-3\cos^2\theta |\Theta_2|^2)\ ,
\label{soft masses2}
\\
m_3^2 &=& m_{3/2}^2 (1-3\cos^2\theta \{|\Theta_3|^2 + |\Theta_A|^2\})\ .
\label{soft masses3}
\end{eqnarray}
Obviously, the masses of the matter fields associated with the
first and second complex planes (\ref{soft masses1},\ref{soft masses2})
are the same as in the multimoduli case of orbifolds without
continuous Wilson lines \cite{Brignole2} 
since the moduli metric has no 
$T_{1,2}$-$A$ mixing. On the other hand,
the soft mass (\ref{soft masses3})
is similar to (\ref{soft masses1},\ref{soft masses2})
with the only difference that the parameter $\Theta_A$
associated to the Wilson line modulus $A$ 
appears\footnote{This is also true for twisted matter $C_p$.
In this case 
${\tilde K}_p=(T_1+T_1^*)^{n^1_p}(T_2+T_2^*)^{n^2_p}(T_3+T_3^*-|A|^2)
^{n^3_p}$, where $n^i_p$ are the modular weights, and therefore
$\log {\tilde K}_p=-\sum_i n^i_p {\hat K}_i$. 
Using (\ref{scalar.mass.square}) we obtain
$m_p^2 = m_{3/2}^2 [1+3\cos^2\theta \{n_p^1|\Theta_1|^2 + 
n_p^2|\Theta_2|^2 + n_p^3(|\Theta_3|^2+|\Theta_A|^2)\}]$.
For particles belonging to the untwisted sector one has 
$n^i_p=-\delta^i_p$ and 
(\ref{soft masses1},\ref{soft masses2},\ref{soft masses3}) are recovered.}. 
Since its contribution is similar to those of the moduli $T_i$
we
expect no changes in the phenomenological properties
obtained in \cite{Brignole2}.
In particular, non-universal scalar masses and tachyons are allowed.
Besides, although there may be scalars with mass bigger than gauginos,
on average the scalars are lighter than gauginos. For example, 
three particles
linked via a renormalizable untwisted Yukawa coupling always fulfill
the sum rule
\begin{eqnarray}
m_1^2 + m_2^2 + m_3^2 &=& |M|^2\ .
\label{sum rule}
\end{eqnarray}
This result is obvious from the property 
$\sum_{\alpha} \Theta_{\alpha}^*\Theta_{\alpha}=1$.
Let us remark that the tree-level gaugino masses for any
compactification scheme are given by
$M_a=\sqrt 3 m_{3/2}\sin \theta e^{-i\gamma_S}$.

Furthermore, since                
$\sum_{j=1,2,3}
\partial_{\alpha} \log {\tilde K}_j
-\partial_{\alpha} \hat{K}=0$ due to property (\ref{kahl5})
and the untwisted Yukawa couplings do not depend on the moduli,
from (\ref{trilinear})
one also obtains 
\begin{eqnarray}
A_{123} &=& -M\ ,
\label{trilinear2}
\end{eqnarray}
as in the case of orbifolds without continuous Wilson lines.
  
If the moduli space contains more than one Wilson line modulus,
as mentioned below (\ref{kahl5}),
we just have to define new parameters $\Theta_{A_k}$ and the
modification introduced in the soft mass (\ref{soft masses3}) 
is straightforwardly
computed using again property (\ref{kahl5})\footnote{
Since twisted moduli appear in the K\"ahler potential (to lowest order)
in the same way as Wilson line moduli \cite{Sabra}, the same
result holds for this case.}, 
$|\Theta_A|^2\rightarrow \sum_k |\Theta_{A_k}|^2$.
A similar situation occurs in the case of the K\"ahler potential
(\ref{kahl2}).

In this case the most general K\"ahler potential including matter
fields can be written as
\begin{eqnarray}
K_3 &=& -\log[(T_3+T_3^*)(U_3+U_3^*)-(B+C^*)(B^*+C)
-(C_3+C'^*_3)(C_3^*+C'_3)]\
\nonumber\\ &&
\simeq {\hat K}_3 + {\tilde K}_3 (C_3+C'^*_3)(C_3^*+C'_3)
\label{kahl6}
\end{eqnarray}
with
\begin{eqnarray}
{\tilde K}_3 &=& [(T_3+T_3^*)(U_3+U_3^*)-(B+C^*)(B^*+C)]^{-1}\ ,
\label{kahl7}
\\
{\hat K}_3 &=& \log {\tilde K}_3\ , 
\label{kahl8}
\end{eqnarray}
Due to property (\ref{kahl8})  
we can use again the same argument than above to 
obtain
\begin{eqnarray}
m_3^2 = {m}'^2_3 = m_{3/2}^2 (1-3\cos^2\theta
\{|\Theta_3|^2 + |\Theta_{U_3}|^2 + |\Theta_B|^2 + |\Theta_C|^2\})\ .
\end{eqnarray}
Therefore the comments below (\ref{soft masses3}) still hold.

On the other hand, 
the term ${\tilde K}_3 C_3 C'_3 + h.c.$ in (\ref{kahl6}) possesses 
the required
structure to solve the $\mu$ problem \cite{GM}. 
The $\mu$ and $B$ parameters are given by
\begin{eqnarray}
\mu
&=&
m_{3/2}
\left(
      1
     -\sqrt3\cos\theta (P\Theta)_{\alpha}^{\dagger}
{\bar \partial}_{\alpha}
\log {\tilde K}_3
\right)
\\
B\mu
&=&
m_{3/2}^2
\left[
      2
     -\sqrt3\cos\theta
      \left\{
             (P\Theta)_{\alpha}
                  {\partial}_{\alpha}\log {\tilde K}_3
            +(P\Theta)_{\alpha}^{\dagger}
{\bar \partial}_{\alpha}\log {\tilde K}_3 
      \right\}
\right.\nonumber\\&&\hspace*{10mm}\left.
     +3\cos^2\theta (P\Theta)_{\alpha}^{\dagger} 
      \left\{
                   2
                  ({\bar \partial}_{\alpha}\log {\tilde K}_3)
                  ({\partial}_{\beta}\log {\tilde K}_3)
            -
  \frac{{\bar \partial}_{\alpha}{\partial}_{\beta}{\tilde K}_3}{
{\tilde K}_3}
      \right\}
(P\Theta)_{\beta}\right]
\end{eqnarray}
Using e.g. (\ref{scalar.mass.square}) one can easily check that 
\begin{eqnarray}
m_{3}^2+|\mu|^2={m}'^2_{3}+|\mu|^2=B\mu\ .
\end{eqnarray}
This is precisely the same result obtained in the case of orbifold
compactifications without including continuous Wilson 
lines \cite{Brignole2}\footnote{The phenomenological 
consequences of this relation in connection with the electroweak
symmetry breaking were
studied in \cite{Rollo}.}.

In summary, we have performed an analysis of the soft SUSY-breaking
terms arising in orbifold compactifications taking into account 
the whole untwisted moduli space. The latter includes not only the
moduli $T_i$ which determine the size and shape of the compact space
but also the continuous Wilson line moduli which are generically present
in any orbifold compactification. 
The final formulae for soft terms are different in general from
those of previous computations where only $T_i$ moduli were considered,
due to the extra contributions of the Wilson line auxiliary fields.
However, the phenomenological properties turn out to be similar
and, in this sense, previous analyses are more general than expected.

\section*{Acknowledgments}
 
We thank D. L\"ust for drawing our attention to the 
issue studied in this paper and for extremely useful discussions.

\noindent HBK is supported by a Ministerio de Educaci\'on y Ciencia grant.
CM is a Brain-Pool Fellow.



\def\MPL #1 #2 #3 {{\em Mod.~Phys.~Lett.}~{\bf#1}\ (#2) #3 }
\def\NPB #1 #2 #3 {{\em Nucl.~Phys.}~{\bf B#1}\ (#2) #3 }
\def\PLB #1 #2 #3 {{\em Phys.~Lett.}~{\bf B#1}\ (#2) #3 }
\def\PR  #1 #2 #3 {{\em Phys.~Rep.}~{\bf#1}\ (#2) #3 }
\def\PRD #1 #2 #3 {{\em Phys.~Rev.}~{\bf D#1}\ (#2) #3 }
\def\PRL #1 #2 #3 {{\em Phys.~Rev.~Lett.}~{\bf#1}\ (#2) #3 }
\def\PTP #1 #2 #3 {{\em Prog.~Theor.~Phys.}~{\bf#1}\ (#2) #3 }
\def\RMP #1 #2 #3 {{\em Rev.~Mod.~Phys.}~{\bf#1}\ (#2) #3 }
\def\ZPC #1 #2 #3 {{\em Z.~Phys.}~{\bf C#1}\ (#2) #3 }

\end{document}